\documentclass{moriond}

\usepackage{amsmath,amssymb}

\newcommand{\ItoP}{$I\!\rightarrow\!P$}
\newcommand{\bell}{B_{\ell}}

\newcommand{\Photo}{}

\begin{document}

\vspace*{4cm}

\title{MAP-BASED ASYMMETRIC BEAM TREATMENT\\[2pt]
{\normalfont\normalsize Application of real-space beam patterns in CMB data analysis}}

\author{ARTEM BASYROV}

\address{Laboratoire AstroParticule et Cosmologie (APC), Universit\'e Paris Cit\'e,
CNRS/IN2P3,\\ 10 rue Alice Domon et L\'eonie Duquet, 75013 Paris, France}

\maketitle\abstracts{
Real telescope beams are asymmetric, breaking the simple $\bell$ factorisation in
harmonic space and coupling modes across $\ell$ and $m$. We develop a map-domain
treatment in which beam convolution is a local, scan-dependent operation applied
directly to the time-ordered data (TOD), accommodating the full beam shape and giving
a direct handle on intensity-to-polarisation (\ItoP{}) leakage. A noiseless
validation against spherical-harmonic convolution reproduces the input temperature
and $B$-mode power spectra up to $\ell\simeq1000$, well beyond the range of primordial
$B$-mode science. We outline how the operation maps onto a sparse linear
\texttt{BeamOperator} for modular map-making.}

\section{Motivation}
Real telescope beams are asymmetric due to optical aberrations, detector
inhomogeneities, and pointing errors~\cite{planck2014beams}. In harmonic space a
symmetric beam reduces to a single window $b_\ell$, whereas an asymmetric beam
requires its full set of coefficients $b_{\ell m}$; the convolution then becomes a
function of beam orientation over the rotation group, with a total-convolution cost
scaling as $\ell_{\max}^{3}\,m_{\max}$~\cite{wandelt2001,febecop}, and harmonic
deconvolution map-making is correspondingly expensive~\cite{keihanen2012}. In the map
domain, beam convolution is instead a local, scan-dependent operation that
accommodates the full beam shape and gives a direct handle on \ItoP{} leakage from
differential beam ellipticity --- a critical $B$-mode systematic far more naturally
treated in pixel space~\cite{febecop}.

\section{Map-domain TOD generation}
We simulate the signal a detector records by convolving a HEALPix sky map with a
pixelised beam pattern along the boresight scan trajectory. For each sample at
pointing $(\theta_t,\phi_t,\psi_t)$,
\begin{equation}
d_t \;=\; \sum_{s} w_s \; m\!\left[\, R(\theta_t,\phi_t,\psi_t)\,\hat{\mathbf{n}}_s \,\right],
\label{eq:tod}
\end{equation}
where $\hat{\mathbf{n}}_s$ and $w_s$ are the direction and weight of beam pixel $s$,
and $R$ is a compound Rodrigues rotation that recenters the beam on the boresight and
applies the polarisation roll $\psi_t$. The algorithm proceeds in three steps:
(i)~prepare the beam pixel map and keep the pixels above a $-25$\,dB power threshold;
(ii)~Rodrigues rotation of the beam pixel map to the local frame $(\theta,\phi,\psi)$;
and (iii)~bilinear interpolation of the sky map onto the beam-pixel positions,
followed by the beam-weighted convolution sum. The implementation is parallelised and
just-in-time (JIT) compiled.

\section{Towards a composable \texttt{BeamOperator}}
The convolution of Eq.~\ref{eq:tod} is linear in the sky, so it can be cast as a
sparse \texttt{BeamOperator} per frequency, detector, and Stokes component, with
off-diagonal blocks carrying polarisation leakage. These can be stacked into a
\texttt{StackedBeamOperator} via block-diagonal/block-column structures, assembling
the full instrument response while preserving sparsity; a future
\textsc{Furax}~\cite{furax} integration would add JAX-native JIT compilation and
autodifferentiation.

\section{Validation against spherical-harmonic convolution}
\begin{figure}[t]
\centering
\begin{minipage}{0.32\linewidth}%
\includegraphics[width=\linewidth]{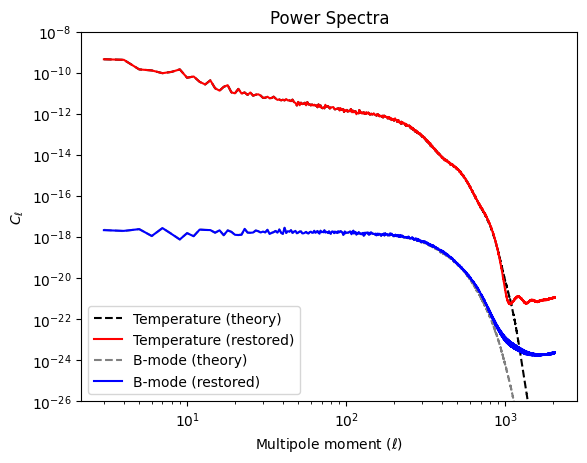}%
\end{minipage}\hfill
\begin{minipage}{0.32\linewidth}%
\includegraphics[width=\linewidth]{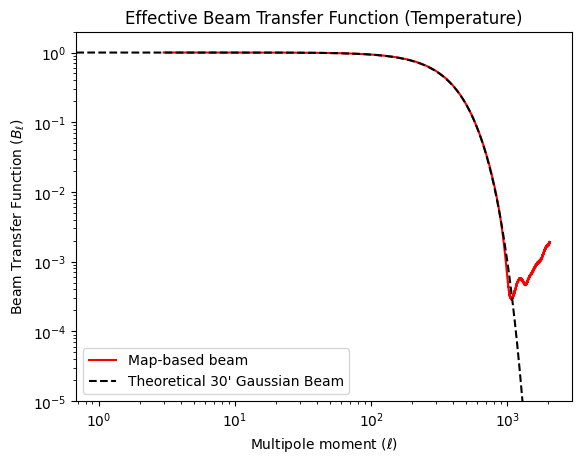}%
\end{minipage}\hfill
\begin{minipage}{0.32\linewidth}%
\includegraphics[width=\linewidth]{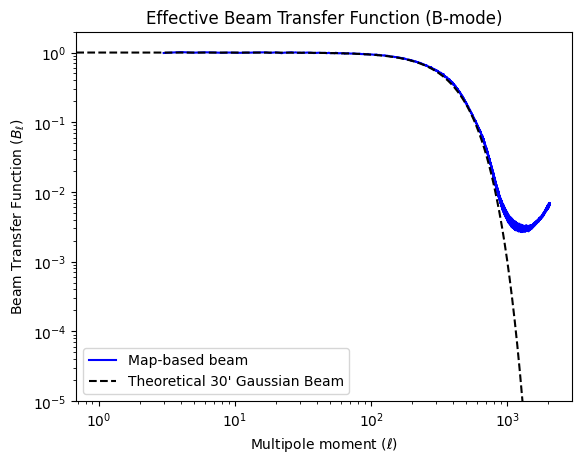}%
\end{minipage}
\caption[]{Reconstructed temperature (red) and $B$-mode (blue) power spectra (left)
and effective beam transfer functions (centre, right), against theory and the
analytic $30'$ Gaussian (dashed). Agreement holds to $\ell\simeq1000$.}
\label{fig:validation}
\end{figure}

To benchmark the method we convolve simulated CMB sky maps ($90$\,GHz, HEALPix
$n_{\rm side}=1024$) with a symmetric $\mathrm{FWHM}=30'$ Gaussian beam on a regular
$1'$ grid, reconstruct maps by map-making, and compare against the same map smoothed
in the spherical-harmonic domain with HEALPix~\cite{healpix}. The reconstructed
temperature and $B$-mode power spectra track the input theory up to $\ell\simeq1000$
(Fig.~\ref{fig:validation}, left), and the effective beam transfer functions agree
with the analytic $30'$ Gaussian over the same range (Fig.~\ref{fig:validation},
centre and right); the residual maps are consistent with noise in both $T$ and $B$.
This is well beyond the range of interest --- the primordial $B$-mode signal lies at
$\ell\lesssim300$ (Simons Observatory~\cite{so_overview,so_forecast}) and
$\ell\lesssim200$ (LiteBIRD~\cite{litebird}).

\section{Conclusions}
Map-domain beam convolution treats asymmetric beams as a local, scan-dependent
operation, giving a natural handle on \ItoP{} leakage. Already before any
operator-based acceleration it reproduces spherical-harmonic convolution to
$\ell\simeq1000$ in both $T$ and $B$, across the full range relevant to Simons
Observatory and LiteBIRD. Casting it as a sparse \texttt{BeamOperator} will make it
directly usable in modern map-making frameworks.

\section*{References}
\bibliography{basyrov}

\end{document}